\documentclass{PoS}

\title{Thermodynamics for SU(2) pure gauge theory using gradient flow}

\ShortTitle{Thermodynamics for SU(2) pure gauge theory using gradient flow}

\author{\speaker{Takehiro Hirakida}\\
        Department of Physics, Graduate School of Science, Kyushu University, Fukuoka 819-0395,
	Japan\\
        E-mail: \email{hirakida@email.phys.kyushu-u.ac.jp}}

\author{Etsuko Itou\\
Department of Mathematics and Physics, Faculty of Science and
Technology, Kochi University, Kochi 780-8520, Japan\\
Department of Physics, and Research and Education Center for Natural
Sciences, Keio University, 4-1-1 Hiyoshi, Yokohama, Kanagawa 223-8521,
Japan\\
Research Center for Nuclear Physics (RCNP), Osaka University, Osaka
567-0047, Japan\\
        E-mail: \email{itou@yukawa.kyoto-u.ac.jp}}

\author{Hiroaki Kouno\\
        Department of Physics, Saga University, Saga 840-8502, Japan\\
        E-mail: \email{kounh@cc.saga-u.ac.jp}}

\abstract{
We present the scale-setting function and the equation of state of the pure SU(2) gauge theory using the gradient flow method.
We propose a reference scale $t_0$ for the SU(2) gauge theory satisfying 
$t^2\langle E \rangle|_{t=t_0} = 0.1$.
This reference value is fixed by a natural scaling-down of the standard $t_0$-scale for the
SU(3) gauge theory based on the perturbative analyses.
We also show the thermodynamic quantities as a function of $T/T_c$, which are derived by the energy-momentum tensor using the small flow time expansion of the gradient flow.
}

\FullConference{The 36th Annual International Symposium on Lattice Field Theory - LATTICE2018\\
		22-28 July, 2018\\
		Michigan State University, East Lansing, Michigan, USA.}

\begin{document}

\section{Introduction}
\label{sec:intro}
The thermodynamic quantities play an important role to understand features of
quantum chromodynamics (QCD).
In the lattice calculation, the thermodynamic quantities for the quenched QCD have been precisely obtained by using several methods; the integral method, the moving-frame method, and the non-equilibrium method.
In the recent study \cite{Asakawa:2013laa}, these quantities are directly calculated from the energy-momentum tensor (EMT) by the gradient flow method.
This method has two advantages.
First is that the statistical uncertainties become smaller with a few hundreds of configurations than that of the other methods.
Actually, we utilize a few hundreds of configurations for the numerical calculation, while the recent study using the integral method \cite{Giudice:2017dor} needs more than 10,000 configurations.
Second is that the wave-function renormalization of the EMT is not necessary in the pure gauge theory.

In this work, we use the gradient flow method for the determination of the scale-setting function and the calculation of the thermodynamic quantities in the pure SU(2) gauge theory.
This theory is a good testing ground for the SU(3) gauge theory, since it has almost the same properties as the SU(3) gauge theory.
On the other hand, the numerical cost is lower than that of the SU(3) gauge theory, since the number of the matrix elements is smaller than that of the SU(3).
This work is the first application of the numerical calculation of the pure SU(2) gauge theory using the gradient flow method.
The details of the numerical results are reported in our paper \cite{Hirakida:2018uoy}.

\section{Gradient flow method}
\label{sec:gf}
The Yang-Mils gradient flow equation~\cite{Luscher:2010iy} is defined by
\begin{eqnarray}
 \label{eq:gf}
 \frac{\partial B_\mu}{\partial t} = D_\nu G_{\mu\nu},~~
 B_\mu(t,x)\bigr|_{t=0} = A_\mu(x),
\end{eqnarray}
where $t$, $A_\mu$, and $B_\mu$ denote a fictitious time (``flow time''),
the quantum gauge field, and the deformed gauge field, respectively.
The field strength, $G_{\mu\nu}$, consists of $B_\mu$.
The solution of this equation defines a transformation of the gauge field toward the stationary
points of the gauge action.
The deformed field can be considered to be the renormalized field by the nonperturbative transformation. 
The flow time can be identified as a typical energy scale of the renormalization.
Due to these properties, the composite operators consisting of $B_\mu$ become UV finite in the positive flow time in the pure gauge theory.

In general, measurement of the EMT using the lattice numerical simulation is difficult,
since the lattice regularization breaks the general covariance.
Here, we calculate the EMT by using the method based on the small flow time expansion of the
Yang-Mills gradient flow~\cite{Luscher:2011bx,Suzuki:2013gza};
\begin{eqnarray}
 \label{eq:EMT1}
 T^R_{\mu\nu}(x) &=& \lim_{t\to 0}\Biggl\{\frac{U_{\mu\nu}(t,x)}{\alpha_U(t)}
 + \frac{\delta_{\mu\nu}}{4\alpha_E(t)}\Bigl[E(t,x) - \langle E(t,x)\rangle_0\Bigr]\Biggr\},\\
 \label{eq:EMT2}
 U_{\mu\nu} &=& G^a_{\mu\rho}G^a_{\nu\rho} - \frac{\delta_{\mu\nu}}{4}G^a_{\rho\sigma}G^a_{\rho\sigma},~
 E = \frac{1}{4}G^a_{\mu\nu}G^a_{\mu\nu},
\end{eqnarray}
where $\langle E \rangle_0$ is the expectation value of $E$ at $T = 0$ and
the coefficients $\alpha_U$ and $\alpha_E$ are calculated in the one-loop order (see Ref.~\cite{Suzuki:2013gza} for the details).
In the finite temperature, the thermodynamic quantities (entropy density ($s$), trace anomaly ($\Delta$),
energy density ($\varepsilon$), and pressure ($P$)) are calculated as following;
\begin{eqnarray}
 \label{eq:entropy-anomaly}
 sT &= \varepsilon + P = T_{11} - T_{44},~
 \Delta = \varepsilon - 3P = - \sum_{\mu=1}^4 T_{\mu\mu}.
\end{eqnarray}

\section{Scale setting}
\label{sec:scale}
  \subsection{Observable and simulation setup}
  \label{ss:observable-and-setup}
  The observable for the scale setting is the dimensionless quantity $t^2E$, where $E$ denotes the action density given in Eq.~(\ref{eq:EMT2}).
  In the perturbative analysis, the leading term of $t^2E$ is proportional to $N_c^2 - 1$.
  The reference scale ($t_0$-scale) for the SU(3) gauge theory is proposed in Ref.~\cite{Luscher:2010iy},
  that satisfies $t^2\langle E(t) \rangle|_{t=t_0} = 0.3$.
  In this work, we introduce the following reference value for the SU(2) gauge theory;
  \begin{eqnarray}
   \label{eq:reference-value}
   t^2\langle E(t)\rangle\Bigr|_{t=t_0} = 0.1.
  \end{eqnarray}
  This value is chosen as an approximate scaling-down of that for the SU(3) gauge theory.

  For the configuration generation, we utilize the Wilson-Plaquette action and set the lattice extent to $N_s^4 = 32$.
  Table~\ref{tab:setup1} shows the lattice coupling constant $\beta$ and the number of configurations for each parameter.
  As a range of the flow time in the gradient flow process, we set $t/a^2 \in [0.00,32.00]$.
  To obtain the reference values ($t_0$), we calculate $E$ using the clover operator as a central analysis.
  The systematic error is estimated  by the difference from the value calculated by the plaquette operator.
   \begin{table}[h]
    \centering
    \caption{Simulation parameter to obtain the scale-setting function.
    We also summarize the results of $t_0/a^2$.}
    \label{tab:setup1}
    \begin{tabular}{|c|cccccc|}
     \hline
     $\beta$ & 2.420 & 2.500 & 2.600 & 2.700 & 2.800 & 2.850 \\ \hline
     \# of Conf. & 100 & 300 & 300 & 300 & 300 & 600 \\ \hline
     $t_0/a^2$ & 1.083(2) & 1.839(3) & 3.522(10) & 6.628(36) & 11.96(12) & 16.95(17)\\ \hline
    \end{tabular}
   \end{table}

   \subsection{Results of the scale setting}
   \label{ss:result1}
   The obtained values of $t_0$ in lattice units are summarized in Table~\ref{tab:setup1}.
   The data of $\ln(t_0/a^2)$ are well interpolated using a quadratic function of $\beta$,
   and we obtain the following best fit function for $\beta \in [2.420,2.850]$;
   \begin{eqnarray}
    \label{eq:scale setting function}
    \ln(t_0/a^2) = 1.285 + 6.409(\beta - 2.600) - 0.7411(\beta - 2.600)^2.
   \end{eqnarray}
   This scale-setting function gives the relation between the lattice spacing $a$ and the temperature for a given $\beta$.
   
\begin{figure}[h]
    \centering
    \includegraphics[height=0.25\textheight,clip]{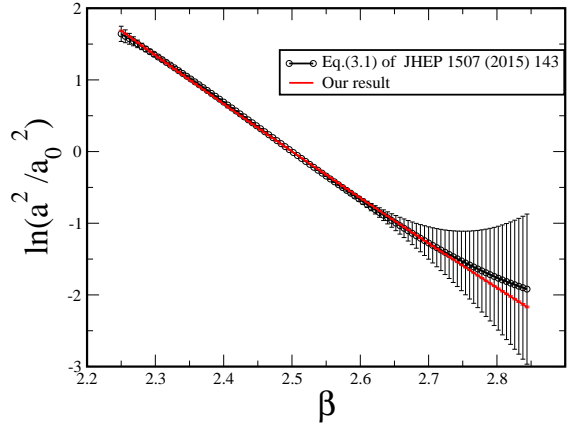}
    \caption{The ratio of the lattice spacing $\ln(a/a_0)$ as a function of $\beta$.}
    \label{comp-latt-space-ratio}
   \end{figure}

   Figure \ref{comp-latt-space-ratio} shows the ratio of the lattice spacing $\ln(a/a_0)$ as a function of $\beta$.
   We take the reference lattice spacing $a_0$ with the value at $\beta=2.500$.
   As a comparison, the previous data, which is obtained from the scale-setting equation in Ref.~\cite{Caselle:2015tza}, is also shown.
   There is a precise agreement between two scale-setting functions within $1$-$\sigma$ errorbar in $\beta \in [2.42,2.60]$, where both functions are available.
   Also, our scale-setting function has the smaller statistical errorbar than that given in Ref.~\cite{Caselle:2015tza} does.
   Our function covers the higher $\beta$ region, which is better to investigate the physics in $T_c \lesssim T$.

   We also compare our reference scale ($\sqrt{8t_0}$) with the Sommer scale ($r_0$) \cite{Sommer:1993ce} and the Necco-Sommer scale ($r_c$) \cite{Necco:2001xg}.
   These reference scales are defined via the dimensionless static quark-antiquark force $r^2F(r)$.
   To estimate the ratios of $\sqrt{8t}$ to $r_0$ and $r_c$, we use the data of $a^2F(r)$ at $\beta = 2.50,2.60$ and $2.70$ shown in Ref.~\cite{Sommer:1993ce}.
   The results in the continuum limit are given by
   \begin{eqnarray}
    \label{eq:t0-r0-rc}
    \sqrt{8t_0}/r_0 = 0.6020(86)(40),~~
    \sqrt{8t_0}/r_c = 1.126(7)(7),
   \end{eqnarray}
   where the first and the second brackets show the statistical and the systematic errors.
   These results imply that our $t_0$-scale is closer to the Necco-Sommer scale.
   If we take $r_0 = 0.50$ [fm], our reference value becomes $\sqrt{8t_0} = 0.3010(43)(20)$ [fm].

\section{Thermodynamics}
\label{sec:thermo}
  \subsection{Procedure and simulation setup}
  \label{ss:prosedure-setup}
  In order to obtain the EMT on the lattice, we perform the following 
  procedures given in Ref.~\cite{Asakawa:2013laa};
  \begin{enumerate}
   \item Generate configurations at $t = 0$.
	 The number of configurations is 200.
	 For the calculation of $U_{\mu\nu}$ and $E$, we use the lattice size $N_s^3 \times N_\tau$ with the fixed aspect ratio $N_s/N_\tau = 4$ for $N_\tau = 6,8,10$ and $12$.
	 The simulation parameters for the finite temperature are shown in Table \ref{tab:beta-T} for each $T/T_c$, which are determined by using Eq.~(\ref{eq:scale setting function}).
	 To find $\langle E \rangle_0$, we also generate the configurations for each $\beta$ listed in Table~\ref{tab:beta-T} with the lattice size $N_s^4 = 32^4$.
   \item Solve the gradient flow equation for each configuration in the fiducial window $2a\le \sqrt{8t} \le N_\tau a/2$ to suppress the discretization and the finite volume effects.
	 Here, we set the lower limit to $\sqrt{8t_{\mathrm{min}}} = 2a$, since we consider the clover operator with a size $2a$.
	 The upper limit is fixed to be $\sqrt{8t_{\mathrm{max}}}=N_\tau a/2$, since the smearing effect by the gradient flow exceeds the temporal lattice size in $\sqrt{8t} > N_\tau a/2$.
   \item Construct the operators, $U_{\mu\nu}$ and $E$, using the deformed gauge field and average them over the gauge configurations at each $t$.
   \item Carry out an extrapolation toward $(a,t) \to (0,0)$, first $a \to 0$ and then $t \to 0$ under the condition in the fiducial window.
	 Here, we carry out both constant and linear extrapolations.
	 In the $a \to 0$ limit, we utilize the constant extrapolation as a central analysis and the linear one to estimate the systematic error.
	 Also in the $t \to 0$ limit, we perform two types of extrapolations, and take the result with the better reduced $\chi^2$.
	 Two extrapolated values in the both limits are consistent with each other within at most $3$-$\sigma$, where $\sigma$ denotes the statistical error.
  \end{enumerate}
  \begin{table}[h]
   \centering
   \caption{Lattice parameters ($\beta$ and $T/T_c$) at finite temperature simulations.}
   \label{tab:beta-T}
   \begin{tabular}{|c||c|c|c|c|}
    \hline
    $T/T_c$ & $N_\tau = 6$ & $N_\tau = 8$ & $N_\tau = 10$ & $N_\tau = 12$ \\ \hline
    0.95 & ---  & 2.50 & 2.57 & 2.62 \\
    0.98 & 2.42 & 2.51 & 2.58 & 2.63 \\
    1.01 & 2.43 & 2.52 & 2.59 & 2.64 \\
    1.04 & 2.44 & 2.53 & 2.60 & 2.65 \\
    1.08 & 2.45 & 2.54 & 2.61 & 2.66 \\
    1.12 & 2.46 & 2.55 & 2.62 & 2.67 \\
    1.28 & 2.50 & 2.59 & 2.66 & 2.72 \\
    1.50 & 2.55 & 2.64 & 2.71 & 2.77 \\
    1.76 & 2.60 & 2.69 & 2.76 & 2.82 \\
    2.07 & 2.65 & 2.74 & 2.81 & ---  \\ \hline    
   \end{tabular}
  \end{table}

  To calculate the thermodynamic quantities, we firstly measure the trace anomaly ($\Delta/T^4$)
  and the entropy density ($s/T^3$), and then compute the energy density
  ($\varepsilon/T^4$) and the pressure ($P/T^4$).

  \subsection{Results of thermodynamics}
  \label{ss:result2}
\begin{figure}[h]
   \centering
   \includegraphics[width=0.95\textwidth,clip]{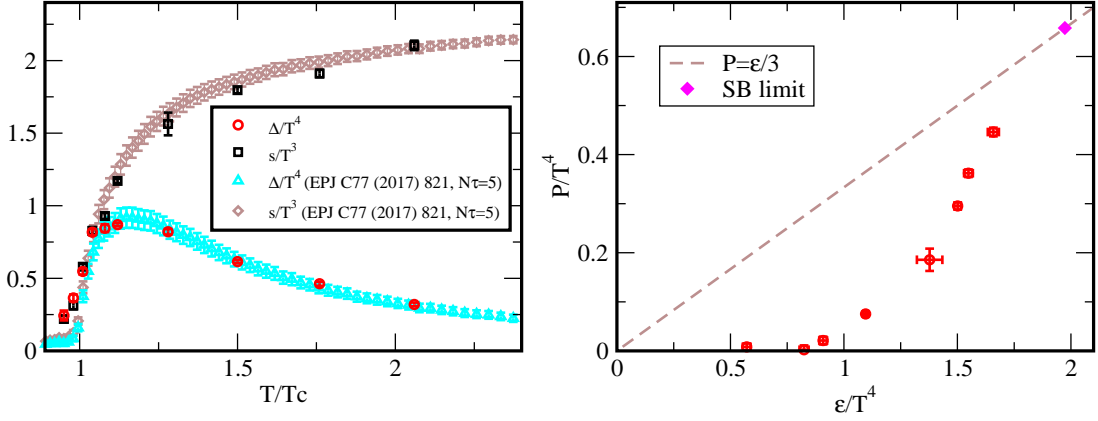}
   \caption{(Left) Results of the trace anomaly (circle) and the entropy density (square) as a function of temperature.
   (Right) The equation of state of our results.
   }
   \label{fig:entropy-anomaly-eos}
  \end{figure}

  We plot $\Delta/T^4$ and $s/T^3$ as a function of $T/T_c$ 
  after taking the double $(a,t)\to (0,0)$ limit in the left panel of Fig.~\ref{fig:entropy-anomaly-eos}.
  As a comparison, $\Delta/T^4$ and $s/T^3$ obtained by the integral method at $N_\tau = 5$ given in Ref.~\cite{Giudice:2017dor} are also shown.
  Our results have the smaller statistical error than the results obtained by the integral method.
  In $T/T_c \ge 1.12$, our data including the systematic errors agree with the results given by the integral method, while there are small discrepancies in the low temperature region.
  These discrepancies may come from the discretization errors in the both methods and the usage of the one-loop approximation in the gradient flow method.

  The right panel of Fig.~\ref{fig:entropy-anomaly-eos} shows the equation of state in $T \ge T_c$, namely the relationship between $\varepsilon/T^4$ and $P/T^4$.
  The linear function, $P = \varepsilon/3$, presents the case with vanishing $\Delta/T^4$, and the diamond symbol denotes the value in the ideal gas (Stefan-Boltzmann (SB)) limit.
  In the high temperature, our result heads toward the point in the SB limit.
  However, the lattice data at $T \simeq 2T_c$ is almost $70$-$80$\% of the value in the SB limit.
  It is an evidence that the state of the two-color ``QGP'' phase around $T \le 2T_c$ cannot be described by the ideal gas model yet.

  Now, let us compare our results with the analytic prediction,
  namely the results of the Hard-Thermal-Loop (HTL) model \cite{Andersen:2010ct}.
  In the HTL analysis, we use the NNLO formula with the renormalization scheme $\mu_\mathrm{HTL} = 2\pi T$.
  The systematic uncertainty of $\mu_\mathrm{HTL}$ is shown as two dashed curves obtained by using $\mu_\mathrm{HTL}/(2\pi T) = 0.5$ and $2.0$.
  The band in the figure shows the uncertainty coming from the error in $T_c/\Lambda_{\overline{\mathrm{MS}}}$ \cite{Fingberg:1992ju}.

\begin{figure}[h]
   \centering
   \includegraphics[width=0.95\textwidth,clip]{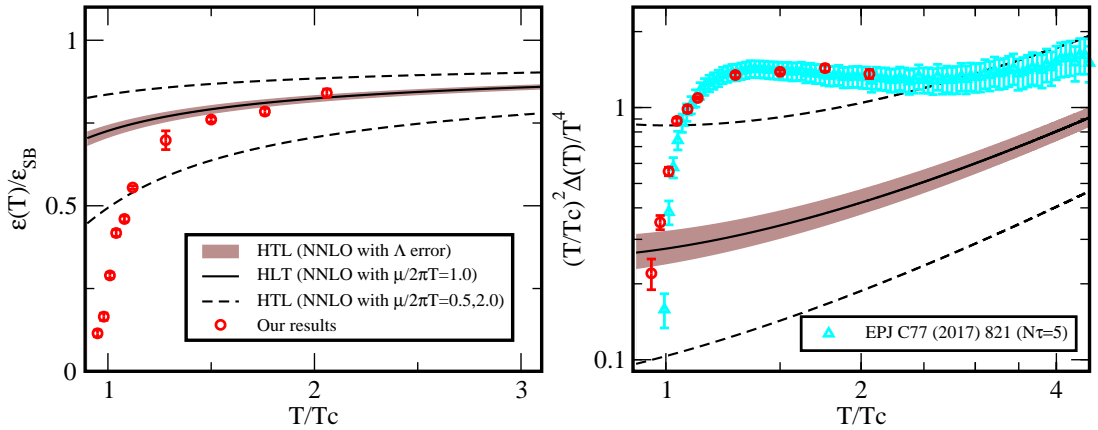}
   \caption{The comparison between our results and the HTL analysis.
   The left panel shows the results of $\varepsilon/\varepsilon_\mathrm{SB}$,
   while the right one presents the results of $(T/T_c)^2\Delta/T^4$.}
   \label{fig:comp-HTL}
  \end{figure}

  The left panel of Fig.~\ref{fig:comp-HTL} shows the comparison of $\varepsilon$ normalized by the value
  in the SB limit between our lattice data and the HTL calculation in $N_c = 2$ case.
  Our result is nicely consistent with the HTL prediction in $T_c \lesssim T$.

  To see the scaling law of the trace anomaly, we also compare the results between the lattice data and the HTL analysis.
  The trace anomaly has a term of $1/T^2$ as a leading correction in the high temperature.
  On the other hand, the nonperturbative logarithmic correction term for $\Delta/T^2$ is predicted by the HTL analysis.
  To see these correction terms, we take the both axes as a logarithmic scale in the right panel of Fig.~\ref{fig:comp-HTL}.
  We also set the horizontal axis to be $(T/T_c)^2\Delta/T^4$.
  Here, as a comparison, the result obtained by the integral method at $N_\tau = 5$ given in Ref.~\cite{Giudice:2017dor} is also shown.
  In $1.3T_c \lesssim T$, the lattice results exhibit almost plateau and approaches to the HTL results.
  We consider that the lattice data become consistent with the HTL and the perturbative analyses in the further high temperature.

\section{Summary}
\label{sec:summary}
In this work, we determine the scale-setting function and the thermodynamic quantities by using the gradient flow method in the case of $N_c = 2$.
For the precise scale-setting of the lattice parameters, we propose a reference value for the SU(2) gauge theory, which satisfies $t^2\langle E \rangle|_{t=t_0}=0.1$.
This value is determined by a natural scaling-down of the reference scale for the SU(3) gauge theory.
We show that our reference scale is suitable to study the thermodynamics near $T_c$ and the statistical error of our scale-setting function is smaller than that of the function given in Ref.~\cite{Caselle:2015tza}.
For the thermodynamic quantities, we calculate the EMT by the small flow-time expansion of the gradient flow method.
The statistical error is smaller than the one given by the integral method, and we find a strong tendency toward the consistency with the HTL prediction in the high temperature.\\
\\
\textbf{Acknowledgments}~~~
We would like to thank M. Yahiro deeply for valuable comments and
discussions.
H.K. is supported in part by a Grant-in-Aid for Scientific Researches No.~17K05446.
E.I. is supported by the Ministry of Education, Culture, Sports,
Science, and Technology (MEXT)-Supported Program for the Strategic
Research Foundation at Private Universities ``Topological Science''
(Grant No.~S1511006)
and in part by the Japan Society for the Promotion of Science (JSPS)
Grant-in-Aid for Scientific Research (KAKENHI) Grant Number 18H01217.
Numerical simulations are performed on xc40 at YITP, Kyoto University, and on SX-ACE at the Research Center 
for Nuclear Physics (RCNP), Osaka University.

\providecommand{\href}[2]{#2}\begingroup\raggedright\endgroup


\begin{thebibliography}{10}

\bibitem{Asakawa:2013laa}
{\scshape FlowQCD} collaboration, M.~Asakawa, T.~Hatsuda, E.~Itou, M.~Kitazawa
  and H.~Suzuki, \href{https://doi.org/10.1103/PhysRevD.90.011501,
  10.1103/PhysRevD.92.059902}{\emph{Phys. Rev.} {\bfseries D90} (2014) 011501}
  [\href{https://arxiv.org/abs/1312.7492}{{\ttfamily 1312.7492}}].

\bibitem{Giudice:2017dor}
P.~Giudice and S.~Piemonte, \href{https://doi.org/10.1140/epjc/s10052-017-5392-6}{\emph{Eur.
  Phys. J.} {\bfseries C77} (2017) 821}
  [\href{https://arxiv.org/abs/1708.01216}{{\ttfamily 1708.01216}}].

\bibitem{Hirakida:2018uoy}
T.~Hirakida, E.~Itou and H.~Kouno, \emph{{Themodynamics for pure SU($2$) gauge
  theory using gradient flow}},
  [\href{https://arxiv.org/abs/1805.07106}{{\ttfamily 1805.07106}}].

\bibitem{Luscher:2010iy}
M.~L{\"u}scher,
  \href{https://doi.org/10.1007/JHEP08(2010)071,
  10.1007/JHEP03(2014)092}{\emph{JHEP} {\bfseries 08} (2010) 071}
  [\href{https://arxiv.org/abs/1006.4518}{{\ttfamily 1006.4518}}].

\bibitem{Luscher:2011bx}
M.~L{\"u}scher and P.~Weisz,
  \href{https://doi.org/10.1007/JHEP02(2011)051}{\emph{JHEP} {\bfseries 02}
  (2011) 051} [\href{https://arxiv.org/abs/1101.0963}{{\ttfamily 1101.0963}}].

\bibitem{Suzuki:2013gza}
H.~Suzuki,
  \href{https://doi.org/10.1093/ptep/ptt059, 10.1093/ptep/ptv094}{\emph{PTEP}
  {\bfseries 2013} (2013) 083B03}
  [\href{https://arxiv.org/abs/1304.0533}{{\ttfamily 1304.0533}}].

\bibitem{Caselle:2015tza}
M.~Caselle, A.~Nada and M.~Panero,
  \href{https://doi.org/10.1007/JHEP11(2017)016,
  10.1007/JHEP07(2015)143}{\emph{JHEP} {\bfseries 07} (2015) 143}
  [\href{https://arxiv.org/abs/1505.01106}{{\ttfamily 1505.01106}}].

\bibitem{Sommer:1993ce}
R.~Sommer, \href{https://doi.org/10.1016/0550-3213(94)90473-1}{\emph{Nucl.
  Phys.} {\bfseries B411} (1994) 839}
  [\href{https://arxiv.org/abs/hep-lat/9310022}{{\ttfamily hep-lat/9310022}}].

\bibitem{Necco:2001xg}
S.~Necco and R.~Sommer,
  \href{https://doi.org/10.1016/S0550-3213(01)00582-X}{\emph{Nucl. Phys.}
  {\bfseries B622} (2002) 328}
  [\href{https://arxiv.org/abs/hep-lat/0108008}{{\ttfamily hep-lat/0108008}}].

\bibitem{Andersen:2010ct}
J.~O. Andersen, M.~Strickland and N.~Su,
  \href{https://doi.org/10.1007/JHEP08(2010)113}{\emph{JHEP} {\bfseries 08}
  (2010) 113} [\href{https://arxiv.org/abs/1005.1603}{{\ttfamily 1005.1603}}].

\bibitem{Fingberg:1992ju}
J.~Fingberg, U.~M. Heller and F.~Karsch,
  \href{https://doi.org/10.1016/0550-3213(93)90682-F}{\emph{Nucl. Phys.}
  {\bfseries B392} (1993) 493}
  [\href{https://arxiv.org/abs/hep-lat/9208012}{{\ttfamily hep-lat/9208012}}].

\end{thebibliography}
\end{document}